\documentstyle[12pt]{article}

\def\thefootnote{\fnsymbol{footnote}}
\def\ref#1{$^{#1)}$}
\def\journal{\topmargin .3in	\oddsidemargin .5in
	\headheight 0pt	\headsep 0pt
	\textwidth 5.625in 
	\textheight 8.25in 
	\marginparwidth 1.5in
	\parindent 2em
	\parskip .5ex plus .1ex		\jot = 1.5ex}
\journal

\catcode`\@=11
\def\marginnote#1{}
\newcount\hour
\newcount\minute
\newtoks\amorpm
\hour=\time\divide\hour by60
\minute=\time{\multiply\hour by60 \global\advance\minute by-\hour}
\edef\standardtime{{\ifnum\hour<12 \global\amorpm={am}%
	\else\global\amorpm={pm}\advance\hour by-12 \fi
	\ifnum\hour=0 \hour=12 \fi
	\number\hour:\ifnum\minute<10 0\fi\number\minute\the\amorpm}}
\edef\militarytime{\number\hour:\ifnum\minute<10 0\fi\number\minute}
\def\draftlabel#1{{\@bsphack\if@filesw {\let\thepage\relax
   \xdef\@gtempa{\write\@auxout{\string
      \newlabel{#1}{{\@currentlabel}{\thepage}}}}}\@gtempa
   \if@nobreak \ifvmode\nobreak\fi\fi\fi\@esphack}
	\gdef\@eqnlabel{#1}}
\def\@eqnlabel{}
\def\@vacuum{}
\def\draftmarginnote#1{\marginpar{\raggedright\scriptsize\tt#1}}
\def\draft{\oddsidemargin -.5truein
	\def\@oddfoot{\sl preliminary draft \hfil
	\rm\thepage\hfil\sl\today\quad\militarytime}
	\let\@evenfoot\@oddfoot	\overfullrule 3pt
	\let\label=\draftlabel
	\let\marginnote=\draftmarginnote
   \def\@eqnnum{(\theequation)\rlap{\kern\marginparsep\tt\@eqnlabel}%
\global\let\@eqnlabel\@vacuum}  }
\def\preprint{\twocolumn\sloppy\flushbottom\parindent 2em
	\leftmargini 2em\leftmarginv .5em\leftmarginvi .5em
	\oddsidemargin -.5in	\evensidemargin -.5in
	\columnsep .4in	\footheight 0pt
	\textwidth 10in	\topmargin  -.4in
	\headheight 12pt \topskip .4in
	\textheight 7.1in \footskip 0pt
	\def\@oddhead{\thepage\hfil\addtocounter{page}{1}\thepage}
	\let\@evenhead\@oddhead	\def\@oddfoot{}	\def\@evenfoot{} }
\def\numberbysection{\@addtoreset{equation}{section}
	\def\theequation{\thesection.\arabic{equation}}}
\def\underline#1{\relax\ifmmode\@@underline#1\else
	$\@@underline{\hbox{#1}}$\relax\fi}
\def\titlepage{\@restonecolfalse\if@twocolumn\@restonecoltrue\onecolumn
     \else \newpage \fi \thispagestyle{empty}\c@page\z@
	\def\thefootnote{\fnsymbol{footnote}} }
\def\endtitlepage{\if@restonecol\twocolumn \else \newpage \fi
	\def\thefootnote{\arabic{footnote}}
	\setcounter{footnote}{0}}  
\catcode`@=12
\relax
\def\figcap{\section*{Figure Captions\markboth
	{FIGURECAPTIONS}{FIGURECAPTIONS}}\list
	{Figure \arabic{enumi}:\hfill}{\settowidth\labelwidth{Figure 999:}
	\leftmargin\labelwidth
	\advance\leftmargin\labelsep\usecounter{enumi}}}
 \relax
\def\tablecap{\section*{Table Captions\markboth
	{TABLECAPTIONS}{TABLECAPTIONS}}\list
	{Table \arabic{enumi}:\hfill}{\settowidth\labelwidth{Table 999:}
	\leftmargin\labelwidth
	\advance\leftmargin\labelsep\usecounter{enumi}}}
 \relax
\def\reflist{\section*{References\markboth
	{REFLIST}{REFLIST}}\list
	{[\arabic{enumi}]\hfill}{\settowidth\labelwidth{[999]}
	\leftmargin\labelwidth
	\advance\leftmargin\labelsep\usecounter{enumi}}}
 \relax
\makeatletter
\newcounter{pubctr}
\def\publist{\@ifnextchar[{\@publist}{\@@publist}}
\def\@publist[#1]{\list
	{[\arabic{pubctr}]\hfill}{\settowidth\labelwidth{[999]}
	\leftmargin\labelwidth
	\advance\leftmargin\labelsep
	\@nmbrlisttrue\def\@listctr{pubctr}
	\setcounter{pubctr}{#1}\addtocounter{pubctr}{-1}}}
\def\@@publist{\list
	{[\arabic{pubctr}]\hfill}{\settowidth\labelwidth{[999]}
	\leftmargin\labelwidth
	\advance\leftmargin\labelsep
	\@nmbrlisttrue\def\@listctr{pubctr}}}
 \relax
\makeatother
\catcode`\@=11
\def\section{\@startsection {section}{1}{0pt}{-3.5ex plus -1ex minus
 -.2ex}{2.3ex plus .2ex}{\raggedright\large\bf}}
\catcode`\@=12
\newskip\humongous \humongous=0pt plus 1000pt minus 1000pt

\newif\ifdtup

\def\oldreffmt#1{\rlap{[#1]} \hbox to 2\parindent{}}

\def\figfmt#1{\rlap{Figure {#1}} \hbox to 1in{}}

\def\beq{\begin{equation}}
\def\eeq{\end{equation}}

\def\bea{\begin{eqnarray}}

\def\eea{\end{eqnarray}}
\catcode`\@=11
\def\eqnarray{\stepcounter{equation}\let\@currentlabel=\theequation
\global\@eqnswtrue
\global\@eqcnt\z@\tabskip\@centering\let\\=\@eqncr
\gdef\@@fix{}\def\eqno##1{\gdef\@@fix{##1}}%
$$\halign to \displaywidth\bgroup\@eqnsel\hskip\@centering
  $\displaystyle\tabskip\z@{##}$&\global\@eqcnt\@ne
  \hskip 2\arraycolsep \hfil${##}$\hfil
  &\global\@eqcnt\tw@ \hskip 2\arraycolsep $\displaystyle\tabskip\z@{##}$\hfil
   \tabskip\@centering&\llap{##}\tabskip\z@\cr}

\def\@@eqncr{\let\@tempa\relax
    \ifcase\@eqcnt \def\@tempa{& & &}\or \def\@tempa{& &}
      \else \def\@tempa{&}\fi
     \@tempa \if@eqnsw\@eqnnum\stepcounter{equation}\else\@@fix\gdef\@@fix{}\fi
     \global\@eqnswtrue\global\@eqcnt\z@\cr}

\catcode`\@=12
\font\tenbifull=cmmib10 
\font\tenbimed=cmmib10 scaled 800
\font\tenbismall=cmmib10 scaled 666
\relax

\begin{document}
\begin{titlepage}
\begin{center}
\today     \hfill    LBL-35682 \\
           \hfill    UCB-PTH-94/11\\
\vskip .5in

{\large \bf Triviality Bounds\\ in \\
the Next to Minimal Supersymmetric Standard Model}
\footnote{This work was supported in part by the Director, Office of
Energy Research, Office of High Energy and Nuclear Physics, Division of
High Energy Physics of the U.S. Department of Energy under Contract
DE-AC03-76SF00098 and in part by the National Science Foundation under
grant PHY90-21139.}

\vskip .5in
Yi-Yen Wu\\[.5in]

{\em  Theoretical Physics Group\\
      Lawrence Berkeley Laboratory\\
      University of California\\
      Berkeley, CA 94720}
\end{center}

\vskip .5in

\newpage
\begin{abstract}
We study the implications of the triviality problem for the
Higgs masses and other relevant parameters
in the Next to Minimal Supersymmetric Standard Model
(NMSSM). By means of triviality, a new way to
constrain parameters is proposed, and therefore
we are able to derive triviality bounds on the
heaviest-Higgs mass, the lightest-Higgs mass,
the soft SUSY-breaking parameters, and the vacuum
expectation value of the Higgs gauge singlet through
a thorough examination of the parameter space.
The triviality upper bound on the lightest-Higgs
mass predicted by NMSSM is indeed larger than
the upper bound predicted by MSSM (the Minimal
Supersymmetric Standard Model).
\end{abstract}
\end{titlepage}
\renewcommand{\thepage}{\roman{page}}
\setcounter{page}{2}
\mbox{ }

\vskip 1in

\begin{center}
{\bf Disclaimer}
\end{center}

\vskip .2in

\begin{scriptsize}
\begin{quotation}
This document was prepared as an account of work sponsored by the United
States Government.  Neither the United States Government nor any agency
thereof, nor The Regents of the University of California, nor any of their
employees, makes any warranty, express or implied, or assumes any legal
liability or responsibility for the accuracy, completeness, or usefulness
of any information, apparatus, product, or process disclosed, or represents
that its use would not infringe privately owned rights.  Reference herein
to any specific commercial products process, or service by its trade name,
trademark, manufacturer, or otherwise, does not necessarily constitute or
imply its endorsement, recommendation, or favoring by the United States
Government or any agency thereof, or The Regents of the University of
California.  The views and opinions of authors expressed herein do not
necessarily state or reflect those of the United States Government or any
agency thereof of The Regents of the University of California and shall
not be used for advertising or product endorsement purposes.
\end{quotation}
\end{scriptsize}

\vskip 2in

\begin{center}
\begin{small}
{\it Lawrence Berkeley Laboratory is an equal opportunity employer.}
\end{small}
\end{center}

\newpage
\renewcommand{\thepage}{\arabic{page}}
\setcounter{page}{1}
\section{Introduction}

\hspace{0.8cm}There have been many studies about the upper bound
on the Higgs mass of the standard model
(or its supersymmetric extension) \cite{ref1,ref2,ref3}.
One of the approaches is based on triviality of the
$\phi^{4}$ theory \cite{ref4}.
Due to triviality, the standard model is inconsistent as a fundamental
theory but is a reasonable effective theory with momentum
cut-off $\Lambda$. Furthermore,
by requiring that $\Lambda$ be larger
than the Higgs mass in order to maintain the consistency
of the standard model as an effective theory, Dashen and
Neuberger were the first
to derive the triviality upper bound (about 800 GeV)
on the Higgs mass in the
minimal standard model \cite{ref4}.
Improvements on this triviality upper bound have
also been made, including the non-perturbative calculations, or
the contributions of gauge couplings and
the top Yukawa coupling \cite{ref5,ref6,ref7,ref8}. So
far, supersymmetry is the only viable framework where
the Higgs scalar is natural \cite{ref9,ref10}.
Therefore, it is important to understand how
this triviality upper bound on the Higgs mass
behaves in the supersymmetric extension of the
standard model, especially the
issue of the lightest-Higgs mass.

The Minimal Supersymmetric Standard Model (MSSM) is the most studied
supersymmetric extension of the standard model.
Another possible extension
is the Next to Minimal Supersymmetric Standard Model (NMSSM)
with two SU(2)$\times$U(1) Higgs doublets and one Higgs singlet
\cite{ref10,ref11,ref12}. The inclusion of the Higgs gauge singlet
in NMSSM provides an explanation to the \mbox{$\mu$ problem} of
MSSM \cite{ref13}. In addition, the existence of the
Higgs singlet is suggested in many superstring models
\cite{ref14,ref15} and
grand  unified supersymmetric models \cite{ref13}.
These features make NMSSM an appealing alternative
to MSSM. An important issue about MSSM is the upper bound on the
lightest-Higgs mass \cite{ref16}. Because there is no guarantee
that there will be
the signal of Higgs particles before
we reach the upper bound predicted by
MSSM, it is definitely interesting to investigate
whether the upper bound on the
lightest-Higgs mass predicted by NMSSM can be larger than
that predicted by
MSSM or not. A pioneering work \cite{ref12} has been done
in this respect.

Similar to the standard model, it is suggested that triviality still
persists in NMSSM when the Higgs couplings are strong. In
short, triviality means that, given the low-energy values of the Higgs
couplings, the Higgs couplings will eventually blow up at
some momentum scale
$\Lambda_{L}$ (the Landau pole) if it is scaled upward,
where $\Lambda_{L}$
is determined essentially by the low-energy values of
the Higgs couplings. The
stronger the low-energy Higgs couplings are, the smaller $\Lambda_{L}$
is. This observation certainly implies an upper bound on the Higgs
mass. In order to establish an upper bound on the lightest-Higgs mass by
means of triviality, one of the possible approaches is to treat NMSSM as
an effective theory with momentum cut-off $\Lambda$, and then require
that the Higgs couplings remain finite beneath the cut-off $\Lambda$.
We will call this approach the Finite
Coupling Constant Formulation (FCCF). In FCCF, based on triviality
associated with the RGE's, the upper bounds on
the low-energy Higgs couplings
can be easily computed once the cut-off $\Lambda$ is specified.
Therefore, the
corresponding upper bound on the lightest-Higgs mass can be
obtained directly
from these triviality upper bounds on the Higgs couplings. The cut-off
$\Lambda$ has to be specified by assuming certain underlying grand
unification scheme ($\Lambda = 10^{15} \sim 10^{17}\,\,\mbox{GeV}$
in most
cases.) There have been several works done in this approach
\cite{ref17,ref18,ref19,ref20}. They all arrived at the same
conclusion
that the upper bound on the lightest-Higgs mass of NMSSM is
indeed larger than
that of MSSM. For example, W.T.A. ter Veldhuis \cite{ref17} reported that
the upper bound on the lightest-Higgs mass of NMSSM will be
\mbox{25 GeV} larger
than that of MSSM if the top-quark mass is \mbox{150 GeV}.

However, according to the spirit of the paper by Dashen and Neuberger
\cite{ref4}, the approach of FCCF is not sufficient and the result is
model-dependent. The formulation of triviality constraints proposed
by Dashen and
Neuberger is based on the requirement that
\mbox{$\Lambda_{L}\,\geq\,m_{HH}$}
($m_{HH}$: the heaviest-Higgs mass) in order to ensure
the consistency of NMSSM as an effective theory with momentum cut-off
$\Lambda\,\leq\,\Lambda_{L}$. We will call this approach the Effective
Theory Consistency Formulation (ETCF). In ETCF, the requirement of
FCCF is always met
because $\Lambda_{L}$ is constructed in such a way that Higgs
couplings blow up at $\Lambda_{L}$. In this sense, ETCF is stronger and
more reasonable than FCCF since ETCF ensures not only the requirement
of FCCF but also the
consistency of NMSSM. ETCF treats $\Lambda_{L}$ as a function of Higgs
couplings and the constraint \mbox{$\Lambda_{L}\,\geq\,m_{HH}$}
represents a constraint on the full parameter space, including the
Higgs couplings and the soft SUSY-breaking parameters because $m_{HH}$
depends on the full parameter space in general. ETCF does
extend the non-trivial implications of triviality to the full
parameter space, whereas the implication of FCCF is limited to the Higgs
couplings. Therefore, ETCF is able to constrain every parameter of the
full parameter space. These constraints will be computed in
\mbox{Section 4}. In addition, there is no need to introduce certain
GUT scheme in ETCF because the cut-off $\Lambda$ is determined
dynamically by the triviality constraint through a thorough search of the
parameter space. In this sense, the approach of ETCF is more general
and model-independent. In conclusion, FCCF is a special case of ETCF.
ETCF provides a more reasonable basis for us to extend the triviality
constraint to the full parameter space. Since ETCF and FCCF are
different in nature, it is worth studying the triviality bounds of
NMSSM based on ETCF.

The purpose of this paper is to describe how to establish triviality
bounds on the full parameter space based on ETCF.
The computation of the NMSSM effective potential in this paper includes
the tree-level contributions only. Therefore, all the triviality bounds
obtained in \mbox{Sections 4 and 5} are tree-level results. However, it
has been pointed out in several works \cite{ref17,ref18,ref20} that
the top and stop loop contributions are quite substantial when the
stop mass is much larger than the top-quark mass. Hence, the present
computations are not very precise. One-loop contributions, including
those of the top and stop, must be included in future computations
in order to make the predictions of the triviality bounds more precise.

In \mbox{Section 2}, the relevant NMSSM lagrangian and renormalization
group
equations are given. Triviality is observed in the case of strong Higgs
couplings, which implies the Landau-pole
behavior of the Higgs couplings. To facilitate the computations of
triviality
bounds, an analytic expression of the Landau pole $\Lambda_{L}$ is also
derived. In \mbox{Section 3}, the parametrization of the NMSSM Higgs
mass spectrum
over the full parameter space is done. The determination of the
full parameter
space is non-trivial because the minimization of the scalar potential
leads to several constraints on the parameters. In \mbox{Section 4},
ETCF is established and the triviality bound is solved through the full
parameter space by requiring \mbox{$\Lambda_{L}\,\geq\,m_{HH}$},
where $\tan\beta=1$ is chosen for the sake of simplicity. Combined with
the present experimental lower bound on the Higgs mass, this analysis
indicates that a very large portion of the parameter space is excluded.
For example, the VEV of the Higgs gauge singlet $v_{3}$ can be
constrained to: \mbox{$0.24M_{W} \leq |v_{3}| \leq 0.749M_{W}$},
where $M_{W}$ is the
mass of W gauge boson. The soft SUSY-breaking parameters are
constrained from
above. Furthermore, the above constraints will become stronger
if the experimental
lower bound on the Higgs mass is raised, which implies a better
understanding
of the correct parameter ranges. In \mbox{Section 5},
an absolute upper bound of
$2.8M_{W}$ on the lightest-Higgs mass is established by
a search through the full
parameter space. This absolute upper bound is beyond the reach of LEP.
\section{Indication of Triviality in NMSSM}

\hspace{0.8cm} The supersymmetric Higgs scalar potential of
the Next to Minimal
Supersymmetric Standard Model (NMSSM) at tree level can be written
as follows \cite{ref10,ref21}:
\begin{eqnarray}
  V&=&{|hN|}^{2}({{\Phi^{\dagger}}_{1}}{\Phi_{1}}+
{{\Phi^{\dagger}}_{2}}{\Phi_{2}})
+{|h{{\Phi^{\dagger}}_{1}}{\Phi_{2}}+\lambda{N^{2}}|}^{2}
+\frac{1}{8}g_{1}^{2}{({{\Phi^{\dagger}}_{1}}
{\Phi_{1}}-{{\Phi^{\dagger}}_{2}}{\Phi_{2}})}^{2} \nonumber \\
& &+\frac{1}{8}g_{2}^{2}[{({{\Phi^{\dagger}}_{1}}{\Phi_{1}}+
{{\Phi^{\dagger}}_{2}}{\Phi_{2}})}^{2}
-4({{\Phi^{\dagger}}_{1}}{\Phi_{2}})({{\Phi^{\dagger}}_{2}}{\Phi_{1}})]
\end{eqnarray}
$\Phi_{1}=({\phi^{\dagger}}_{1},{\phi^{0}}_{1})$ and
$\Phi_{2}=({\phi^{\dagger}}_{2},{\phi^{0}}_{2})$
are two SU(2)$\times$U(1)
Higgs doublets, and $N$ is a complex singlet. $h$ and $\lambda$ are the
Higgs couplings. It is assumed that only $\phi_{1}^{0}$, $\phi_{2}^{0}$
and $N$ acquire non-trivial VEV's
$v_{1}$, $v_{2}$ and $v_{3}$ respectively. The scalar-quarks and
scalar-leptons do not acquire VEV's, and we can ignore
their contributions to the scalar potential when studying the Higgs mass
spectrum. Note that the superpotential corresponding to (1) does not
contain
linear and bilinear terms \cite{ref10} because these terms lead to
naturalness
problems. Besides, these terms do not appear in a large class of
superstring-inspired models. As for the soft SUSY-breaking terms, a
particular $V_{soft}$ is chosen:
\begin{eqnarray}
  V_{soft}&=&m_{1}^{2}{{\Phi^{\dagger}}_{1}}
{\Phi_{1}}+m_{2}^{2}{{\Phi^{\dagger}}_{2}}{\Phi_{2}}
-m_{12}^{2}{{\Phi^{\dagger}}_{1}}{\Phi_{2}}-
{m^{\ast}}_{12}^{2}{{\Phi^{\dagger}}_{2}}{\Phi_{1}}
\end{eqnarray}
in order to have predictive power. However, this particular choice of
$V_{soft}$
does not destroy the generality of the conclusions obtained
in this paper.
The most general $V_{soft}$ will be considered in the last section.

We assume three generations of quarks and leptons together with their
supersymmetric partners. As for the renormalization group equations
relevant
to the Higgs couplings, all the Yukawa couplings are neglected except
for the top Yukawa coupling $f_{t}$. The relevant one-loop RGE's
of NMSSM are given in \cite{ref15,ref21}:
\begin{eqnarray}
  8{\pi^{2}}\frac{d\,{g_{1}^{2}}}{dt}&=&11{g_{1}^{4}} \\
  8{\pi^{2}}\frac{d\,{g_{2}^{2}}}{dt}&=&g_{2}^{4} \\
  8{\pi^{2}}\frac{d\,{g_{3}^{2}}}{dt}&=&-3{g_{3}^{4}} \\
  8{\pi^{2}}\frac{d\,{f_{t}^{2}}}{dt}&=&f_{t}^{2}(6{f_{t}^{2}}+
h^{2}-\frac{13}{9}{g_{1}^{2}}-3{g_{2}^{2}}
                                -\frac{16}{3}{g_{3}^{2}}) \\
  8{\pi^{2}}\frac{d\,{h^{2}}}{dt}&=&h^{2}(4{h^{2}}+
2{\lambda^{2}}+3{f_{t}^{2}}-g_{1}^{2}-3{g_{2}^{2}}) \\
  8{\pi^{2}}
\frac{d\,{\lambda^{2}}}{dt}&=&6{\lambda^{2}}({\lambda^{2}}+h^{2})
\end{eqnarray}
where $g_{1}$, $g_{2}$ and $g_{3}$ are the gauge couplings
associated with
SU(3), SU(2) and U(1) gauge groups respectively.
The parameter t is defined as:
\begin{eqnarray}
 \mbox{t}&=&\frac{1}{2}\,{\ln (\frac{-q^{2}}{M_{W}^{2}})}
\end{eqnarray}
where $q^{2}$ is the space-like effective square of the momentum
at which
these couplings are defined. At t=0, the gauge couplings can be
determined
from the experimentally derived inputs \cite{ref22}:
\begin{eqnarray}
  g_{1}^{2}&=&0.126,\;\;g_{2}^{2}=0.446,\;\;g_{3}^{2}=1.257
\end{eqnarray}

To understand the triviality problem, strong Higgs couplings are
assumed, and therefore the top Yukawa coupling $f_{t}$ and
all the gauge couplings can be ignored. This is not an
unreasonable assumption considering the following
observations: The RGE's \mbox{(3)-(8)}
have been studied numerically by Babu and Ma
\cite{ref21}. Their results indicate
that gauge couplings are negligible if the ratio
$\frac{h^{2}}{g_{1}^{2}}\geq 5$ or
$\frac{\lambda^{2}}{g_{1}^{2}}\geq 2$ holds,
where $\tan\beta$=1 and the top-quark mass $m_{t}$=40$\sim$400 GeV.
(See Fig.1 in \cite{ref21} for a more precise description.)
Therefore, it's certainly reasonable to expect that
\mbox{$\frac{h^{2}}{g_{1}^{2}}\geq 5$} or
\mbox{$\frac{\lambda^{2}}{g_{1}^{2}}\geq 2$} holds
in the case of strong Higgs couplings.
For the sake of self-consistency, the assumption of negligible gauge
couplings in the case of strong Higgs couplings will be verified
\mbox{$a\; posteriori$} by the numerical results of
\mbox{Sections 4 and 5}. The
assumption of negligible Yukawa coupling $f_{t}$ is a little obscure.
There
has been some numerical evidence in the standard model \cite{ref5}
that the determination
of triviality bounds on the Higgs mass is insensitive to
the top-quark mass
$m_{t}$ if $\mbox{$m_{t}\leq$ 200 GeV}$, and it may still
be true in NMSSM.
The assumption of negligible $f_{t}$ will be checked by
the computations of \mbox{Sections 4 and 5}, and
it turns out that $f_{t}$
is important only in certain extreme situations. A detailed discussion
will be given in \mbox{Section 4}.

In the case of strong Higgs couplings
with negligible $f_{t}$ and $g_{i}$,
the one-loop RGE's for the Higgs couplings $h$ and $\lambda$ are:
\begin{eqnarray}
  8{\pi^{2}}\frac{d\,{h^{2}}}{dt}&=&h^{2}(4{h^{2}}+2{\lambda^{2}}) \\
  8{\pi^{2}}
\frac{d\,{\lambda^{2}}}{dt}&=&6{\lambda^{2}}({\lambda^{2}}+h^{2})
\end{eqnarray}
There is only one fixed point, the infrared stable fixed point at
$h^{2}$=0, $\lambda^{2}$=0. Therefore, similar to the Landau pole
\cite{ref3,ref10} of the
pure $\phi^{4}$ theory, $h^{2}$(t) and $\lambda^{2}$(t) diverge at some
finite t=$\mbox{t}_{L}$ (the Landau pole) unless the Higgs couplings
vanish.
Triviality is clearly indicated at one-loop.
Notice that the above conclusion
is still valid even if the top Yukawa coupling $f_{t}$ is included.
The general solution of the Landau pole $\mbox{t}_{L}$ can be found by
means of the method of integrating factor \cite{ref23}:
\begin{eqnarray}
  \mbox{t}_{L}&=&\frac{2{\pi^{2}}
{\lambda_{0}^{-\frac{2}{3}}}\sqrt{1+C{\lambda_{0}^{-\frac{4}{3}}}}}{C}
-\frac{2{\pi^{2}}\ln
(\sqrt{1+C{\lambda_{0}^{-\frac{4}{3}}}}
+\sqrt{C{\lambda_{0}^{-\frac{4}{3}}}}\;)}
          {C^{\frac{3}{2}}} \\
  C&=&{h_{0}^{4}}\,{\lambda_{0}^{-\frac{8}{3}}}+2\,{h_{0}^{2}}\,
{\lambda_{0}^{-\frac{2}{3}}} \\
  h_{0}&=&h(\mbox{t}=0),\;\;\lambda_{0}=\lambda(\mbox{t}=0),\;\;
\Lambda_{L} \equiv M_{W}\cdot{\mbox{exp}(\mbox{t}_{L})}
\end{eqnarray}
where $\Lambda_{L}$ is the momentum corresponding to the Landau pole
$\mbox{t}_{L}$.
\mbox{(13)-(15)} will be useful to the computations of triviality bounds
in \mbox{Section 4}.

Notice that the treatment of triviality here
is of perturbative nature. Although the problem of triviality should
be of
non-perturbative nature, several non-perturbative numerical
simulations have
been performed \cite{ref24,ref25}
and the results indicated that the renormalized perturbative
calculation gives essentially the correct triviality upper bound
on the Higgs
mass. This observation may justify our approach as a first
approximation.
\section{Parameter Space and Higgs Mass Spectrum}

\hspace{0.8cm}Consider the tree-level scalar potential
$V'$=$V$+$V_{soft}$, and
the relevant parameters are {\em h}, $\lambda$, $v_{1}$,
$v_{2}$, $v_{3}$, $m_{1}^{2}$, $m_{2}^{2}$, $m_{12}^{2}$. Without loss of
generality, our convention is to take {\em h}, $\lambda$, $m_{1}^{2}$,
$m_{2}^{2}$ to be
real, and $v_{1}$, $v_{2}$, $v_{3}$, $m_{12}^{2}$ to be complex, i.e.,
$v_{1}={\tilde{v}}_{1}\,e^{i{{\phi}_{1}}}$,
$v_{2}={\tilde{v}}_{2}\,e^{i{{\phi}_{2}}}$,
$v_{3}={\tilde{v}}_{3}\,e^{i{{\phi}_{3}}}$,
$m_{12}^{2}={\tilde{m}}_{12}^{2}\,e^{i{{\phi}_{m}}}$.
The minimization of the scalar potential $V'$ leads to three
complex vacuum constraints on these parameters:
\begin{eqnarray}
  h^{2}({|v_{1}|}^{2}+{|v_{2}|}^{2})v_{3}
+2\lambda(h{v^{\ast}_{1}}v_{2}+\lambda{v_{3}^{2}}){v^{\ast}_{3}}&=&0 \\
\frac{1}{4}(g_{1}^{2}+g_{2}^{2})({|v_{1}|}^{2}-{|v_{2}|}^{2})
+h^{2}({|v_{2}|}^{2}+{|v_{3}|}^{2})
+m_{1}^{2}&=&\frac{v_{2}}{v_{1}}
(m_{12}^{2}-h\lambda{{v^{\ast}}_{3}^{2}}) \\
\frac{1}{4}(g_{1}^{2}+g_{2}^{2})({|v_{2}|}^{2}-{|v_{1}|}^{2})
+h^{2}({|v_{1}|}^{2}+{|v_{3}|}^{2})+m_{2}^{2}&=&
\frac{v_{1}}{v_{2}}({m^{\ast}}_{12}^{2}-h\lambda{v_{3}^{2}})
\end{eqnarray}
In addition, one has the following physical constraint:
\begin{eqnarray}
  M_{W}^{2}&=&\frac{1}{2}g_{2}^{2}({|v_{1}|}^{2}+{|v_{2}|}^{2})
\end{eqnarray}
Imaginary parts of the constraints (16)-(18) fix the phases
among the complex
parameters $v_{1},\;v_{2},\;v_{3},\;m_{12}^{2}$, and (19) reduces
(${\tilde{v}}_{1},\;{\tilde{v}}_{2}$) to a single parameter
$\tan\beta=\frac{{\tilde{v}}_{2}}{{\tilde{v}}_{1}}$. Furthermore,
($h,\;\lambda,\;{\tilde{v}}_{3}$) can be expressed in terms of other
parameters by means of the real parts of the constraints (16)-(18). The
parameter space under study is then defined as the set of parameters
\mbox{($\phi,\;\tan\beta,\;m_{1}^{2},\;m_{2}^{2},\;{\tilde{m}}_{12}^{2}$)},
where
\begin{equation}
  0\,\leq\,\phi\,<\,4\pi,\;\; -\infty\,<\,\tan\beta,\;m_{1}^{2},\;
m_{2}^{2},\;{\tilde{m}}_{12}^{2}\,<\,\infty
\end{equation}
The other parameters can be expressed in terms of (20):
\begin{eqnarray}
  {\tilde{v}}_{3}&=&\frac{M_{W}}{g_{2}}\sqrt{\frac{2\,A\,B\,
\tan\beta-B^{2}(1+{\tan^{2}\beta})}{A^{2}(1+{\tan^{2}\beta})}}
\nonumber \\
  A&=&{\tilde{m}}_{12}^{2}(\tan\beta-\frac{1}{\tan\beta})
-m_{1}^{2}+m_{2}^{2}
      +M_{W}^{2}(1+\frac{g_{1}^{2}}{g_{2}^{2}})
\frac{{\tan^{2}\beta}-1}{{\tan^{2}\beta}+1}  \nonumber \\
  B&=&\frac{1}{\tan\beta}\{m_{1}^{2}+\frac{1}{2}
(1+\frac{g_{1}^{2}}{g_{2}^{2}})M_{W}^{2}\}
      -\tan\beta\{m_{2}^{2}+\frac{1}{2}
(1+\frac{g_{1}^{2}}{g_{2}^{2}})M_{W}^{2}\} \\
  v_{1}&=&\frac{M_{W}}{g_{2}}\,\frac{\sqrt{2}}
{\sqrt{{\tan^{2}\beta}+1}},\;
 v_{2}=\frac{M_{W}}{g_{2}}\,\frac{\sqrt{2}\,\tan\beta}
{\sqrt{{\tan^{2}\beta}+1}}\,{e^{i\phi}},\;
 v_{3}={\tilde{v}}_{3}\,e^{i\frac{\phi}{2}} \\
  m_{12}^{2}&=&{\tilde{m}}_{12}^{2}\,{e^{-i\phi}} \\
  h^{2}&=&\frac{{\tilde{m}}_{12}^{2}\,\tan\beta-m_{1}^{2}
          +\frac{M_{W}^{2}}{2}(1+\frac{g_{1}^{2}}{g_{2}^{2}})
\frac{{\tan^{2}\beta}-1}{{\tan^{2}\beta}+1}}
          {{{\tilde{v}}_{3}}^{2}+\tan\beta\,\frac{M_{W}^{2}}{g_{2}^{2}}
          (\frac{\tan\beta}{{\tan^{2}\beta}+1}\,\pm\,
          \sqrt{\frac{{\tan^{2}\beta}}{{(\,{\tan^{2}\beta}+1\,)}^{2}}\,
-\,\frac{g_{2}^{2}\,{{{\tilde{v}}_{3}}^{2}}}{M_{W}^{2}}}\;)} \\
  \lambda&=&h\,\frac{M_{W}^{2}}{g_{2}^{2}\,{{{\tilde{v}}_{3}}^{2}}}
            (-\frac{\tan\beta}{{\tan^{2}\beta}+1}\,\pm\,
            \sqrt{\frac{{\tan^{2}\beta}}{{(\,{\tan^{2}\beta}+1\,)}^{2}}\,
-\,\frac{g_{2}^{2}\,{{{\tilde{v}}_{3}}^{2}}}{M_{W}^{2}}}\;)
\end{eqnarray}
\begin{equation}
  {\tilde{v}}_{3},\; h,\; \mbox{and}\; \lambda\; \mbox{are real.}
\end{equation}
$''\pm''$ in (24) and (25) indicates ($h,\;\lambda$) has two solutions,
where
the solution with $''+''$ is denoted as ($h_{A},\;\lambda_{A}$)
and the solution
with $''-''$ is denoted as ($h_{B},\;\lambda_{B}$).

(20)-(26) specify the full parameter space. (26) is non-trivial because
square roots are involved in (21), (24) and (25). Together with (25), the
reality condition of $\lambda$ leads to:
\begin{equation}
  {\tilde{v}}_{3}\,\leq\,
\frac{|\tan\beta|}{g_{2}({\tan^{2}\beta}+1)}M_{W}
\end{equation}
$\frac{|\tan\beta|}{tan^{2}\beta\,+\,1}$ has its maximum=$\frac{1}{2}$ at
$\tan\beta=\pm 1$. Therefore, we are able to establish
an absolute upper bound
on ${\tilde{v}}_{3}$:
\begin{equation}
  {\tilde{v}}_{3}\,\leq\,\frac{1}{2\,g_{2}}M_{W}
\end{equation}
Given $M_{W}\,=\,80\,\mbox{GeV}$ and $g_{2}^{2}\,=\,0.446$ from
(10), this
absolute upper bound on the magnitude of $v_{3}$ is \mbox{60 GeV}. As
${\tilde{v}}_{3}\,\rightarrow\,\frac{M_{W}}{2{g_{2}}}$, (27) implies
$|\tan\beta|\,\rightarrow\,1$. Therefore, in NMSSM, $\tan\beta$ can be
constrained by means of $|v_{3}|$, and vice versa.

For the sake of simplicity, we choose $\tan\beta=1$
when studying (20)-(26).
When $\tan\beta=1$, ${\tilde{v}}_{3}$ becomes a free parameter, and
$m_{1}^{2}=m_{2}^{2}$ is required by the minimization of the scalar
potential
$V'$: (17) and (18). On the whole, the number of free parameters is
unchanged. The full parameter space ($\tan\beta=1$) is then defined as
the set of parameters
\mbox{($\phi,\;{\tilde{v}}_{3},\;m_{1}^{2}=
m_{2}^{2},\;{\tilde{m}}_{12}^{2}$)}
plus the following constraints:
\begin{equation}
  0\,\leq\,\phi\,<\,4\pi,\;\; 0\,<\,
{\tilde{v}}_{3}\,\leq\,\frac{1}{2\,g_{2}}M_{W}
\end{equation}
\begin{equation}
  -\infty\,<\,m_{1}^{2}=m_{2}^{2}\,<\,{\tilde{m}}_{12}^{2}\,<\,\infty
\end{equation}
There is no essential change to the expressions of the other parameters
except for {\em h} and $\lambda$:
\begin{eqnarray}
  h^{2}&=&\frac{{\tilde{m}}_{12}^{2}-m_{1}^{2}}
{{\tilde{v}}_{3}^{2}+\frac{M_{W}^{2}}{g_{2}^{2}}
          (\frac{1}{2}\,\pm\,\sqrt{\frac{1}{4}\,-\,\frac{g_{2}^{2}\,
{\tilde{v}}_{3}^{2}}{M_{W}^{2}}}\;)} \\
  \lambda&=&h\,\frac{M_{W}^{2}}{g_{2}^{2}\,{{{\tilde{v}}_{3}}^{2}}}
            (-\frac{1}{2}\,\pm\,\sqrt{\frac{1}{4}\,-\,
\frac{g_{2}^{2}{\tilde{v}}_{3}^{2}}{M_{W}^{2}}}\;)
\end{eqnarray}
(29)-(32) specify the full parameter space and will be studied later.
Notice that (31) and (32) imply:
\begin{equation}
  h^{2},\;\lambda^{2}\;\;\propto\;\;({\tilde{m}}_{12}^{2}-m_{1}^{2})
\end{equation}
and $m_{1}^{2}=m_{2}^{2}<{\tilde{m}}_{12}^{2}$ in (30) is
the consequence of
the reality condition of $h$.

Based on the parameter space described in (20) or (29),
it is trivial to work
out the Higgs squared-mass spectrum from the tree-level potential $V'$.
In general,
$\{\phi_{1}^{0},\;\phi_{2}^{0},\;N\}$ does not mix with
$\{\phi_{1}^{\dagger},\;\phi_{2}^{\dagger}\}$. The squared-mass matrix
$[M_{n}^{2}]$ for $\{\phi_{1}^{0},\;\phi_{2}^{0},\;N\}$ is a 6$\times$6
matrix, and the squared-mass matrix $[M_{c}^{2}]$ for
$\{\phi_{1}^{\dagger},\;\phi_{2}^{\dagger}\}$ is a 4$\times$4 matrix.
As expected, $[M_{n}^{2}]$ contains five neutral Higgs bosons and
one massless particle.
$[M_{c}^{2}]$ contains two charged Higgs bosons and two massless
particles.
The detailed expressions of $[M_{n}^{2}]$ and $[M_{c}^{2}]$ have
been given in \cite{ref20} and won't be repeated here. However,
there are
several symmetries of $[M_{n}^{2}]$ and $[M_{c}^{2}]$:
\begin{eqnarray}
  [M_{n}^{2}]\;\mbox{and} \; [M_{c}^{2}]\; \mbox{are periodic in}\;
\phi,\; \mbox{of periodicity}\; \pi.
\end{eqnarray}
In addition, [$M_{n}^{2}$] and [$M_{c}^{2}$] are invariant under
two discrete symmetries on the parameter
space:
\begin{equation}
  h\,\rightarrow\,-h,\;\; \lambda\,\rightarrow\,\lambda,\;\;
\tan\beta\,\rightarrow\,-\tan\beta,\;\;
  m_{12}^{2}\,\rightarrow\,-m_{12}^{2}
\end{equation}
\begin{equation}
  h\,\rightarrow\,-h,\;\; \lambda\,\rightarrow\,-\lambda
\end{equation}
As for the Higgs squared-mass spectrum, we are interested in these
two quantities: $m_{HH}$ (the heaviest-Higgs mass)
and $m_{LH}$ (the lightest-Higgs mass). In \mbox{Sections 4 and 5}, the
triviality bounds $B_{HH}$ (the upper bound on $m_{HH}$) and
$B_{LH}$ (the upper bound on $m_{LH}$) will be derived. In general,
$m_{HH}$ and $m_{LH}$ have to be computed from $[M_{n}^{2}]$ numerically,
and there are no simple analytic expressions. However,
the understanding of
the dependence of the lightest-Higgs mass $m_{LH}$ on the Higgs coupling
$h$ will be very useful in establishing the triviality bounds.
Using the fact that the smallest eigenvalue is smaller than the smallest
diagonal term and choosing an appropriate basis for $[M_{n}^{2}]$,
Bin\'{e}truy and Savoy \cite{ref19} have derived
an upper bound on $m_{LH}$
\mbox{($\tan\beta\,=\,1$)}:
\begin{equation}
B_{LH}^{(BS)}=\frac{h}{\sqrt{2}}v\,\geq\,m_{LH}, \;\;\;\;\;\;
v\approx 250\,\mbox{GeV}
\end{equation}
Notice that (37) is the result of tree-level computations. This upper
bound is denoted as $B_{LH}^{(BS)}$ in order to be distinguished from
$B_{LH}$ (the triviality upper bound on $m_{LH}$). Therefore, the
dependence of $m_{LH}$ on $h$ can be understood as:
\mbox{$m_{LH}\leq B_{LH}^{(BS)}\,\propto\,h$},
where $B_{LH}^{(BS)}$ is proportional
to the Higgs coupling $h$. Combined with the present experimental
lower bound on
the Higgs mass, \mbox{$m_{LH}\leq B_{LH}^{(BS)}=\frac{h}{\sqrt{2}}v$}
implies that
$h$ must be bounded from below. Notice that $B_{LH}^{(BS)}$ is
introduced for
illustrative purpose only. In practice, the precise determination
of this lower
bound on $h$ is made by solving $m_{LH}$ from $[M_{n}]$ and requiring
\mbox{$\;m_{LH}\geq\,$the experimental lower bound}. By means of
triviality, it will be
argued in \mbox{Section 4} that $h$ decreases with the
soft SUSY-breaking parameter
$m_{1}$, which enables us to establish an upper bound on the
soft SUSY-breaking
parameter $m_{1}$ based on the lower bound on $h$. Details will be
given in
\mbox{Section 4}.
\section{Constraints on the Higgs Mass and \mbox{the Soft
SUSY-Breaking Parameters}}

\hspace{0.8cm}Based on the parameter space
($\phi,\;{\tilde{v}}_{3},\;m_{1}^{2}=m_{2}^{2},\;{\tilde{m}}_{12}^{2}$)
specified by
(29)-(32), \mbox{$\Lambda_{L}\,\geq\,m_{HH}$} ($m_{HH}$: the
heaviest-Higgs mass)
is required by ETCF, and $B_{HH}$ (the triviality
upper bound on the heaviest-Higgs mass $m_{HH}$) is established
by \mbox{$\Lambda_{L}=m_{HH}\equiv B_{HH}$}, where $\Lambda_{L}$ is
defined in (13)-(15) and $m_{HH}$ is computed from $[M_{n}^{2}]$
numerically. Geometrically, the triviality upper
bound $B_{HH}$ defines a
surface in the parameter space by means of
\mbox{$\Lambda_{L}=m_{HH}$}, and our
convention is to parametrize this triviality surface in terms of
\mbox{($\phi,\;{\tilde{v}}_{3},\;m_{1}^{2}=m_{2}^{2}$)},
where ${\tilde{m}}_{12}^{2}$
depends on \mbox{($\phi,\;{\tilde{v}}_{3},\;m_{1}^{2}=m_{2}^{2}$)}
through \mbox{$\Lambda_{L}=m_{HH}$}. At any point
($\phi,\;{\tilde{v}}_{3},\;m_{1}^{2}=m_{2}^{2}$) of this triviality
surface, the seven non-zero eigenvalues of $[M_{n}]$ and $[M_{c}]$
define the seven triviality upper bounds on the seven physical
Higgs masses
respectively. For example, the triviality upper bound on $m_{LH}$
($m_{LH}$: the lightest-Higgs mass)
is defined as the $m_{LH}$ evaluated on the triviality surface.
Therefore, care should be taken in distinguishing $m_{LH}$ from the
triviality
upper bound on $m_{LH}$. Next, we will study the general features of the
triviality surface.

A typical example is chosen as:
$(h,\;\lambda)=(h_{A},\;\lambda_{A})$, $\phi=0$, ${\tilde{v}}_{3}
=0.7\,{M_{W}}$. Its triviality surface is computed, and $B_{HH}$
versus $m_{1}^{2}\,(=m_{2}^{2})$ is plotted in Fig.1.
Fig.1 corresponds to
a line on the triviality surface. Several universal features of
Fig.1 are important. First, when $m_{1}^{2}$ is small, the soft
SUSY-breaking
terms are not important and therefore the determination of
$B_{HH}$ is insensitive to $m_{1}^{2}$.
Second, the curve in Fig.1 ends at
$m_{1}^{2}\,\approx\,-M_{W}^{2}$ because the squared-mass matrix
$[M_{n}^{2}]$ or $[M_{c}^{2}]$ will develop negative eigenvalues if
$m_{1}^{2}$ becomes too negative. Together with (30), this observation
indicates
that $m_{1}^{2},\;m_{2}^{2},\;{\tilde{m}}_{12}^{2}$ are bounded
from below.
Since nothing interesting happens when $m_{1}^{2}\,<\,0$, we will assume
$0\,\leq\,m_{1}=m_{2}\,<\,{\tilde{m}}_{12}$ from now on.

The last but most important universal feature of Fig.1 is:
When $m_{1}^{2}$
is large, the linear relation \mbox{$B_{HH}\,=\,\sqrt{2}\,m_{1}$}
is a good approximation. This observation can be understood as follows.
(13) implies that, on the triviality surface,
\mbox{$h^{2}\;\mbox{and}\;\lambda^{2}\;\rightarrow\;0$} if
\mbox{$B_{HH}\,(=\Lambda_{L})\;\rightarrow\;\infty$}.
The structure of $[M_{n}^{2}]$ also implies
\mbox{$B_{HH}\;\rightarrow\;\infty$} if
\mbox{$m_{1}=m_{2}\;\rightarrow\;\infty$}.
The above two observations lead to:
\begin{equation}
  \mbox{On the triviality surface,}\;\; h^{2}\;\mbox{and}\;
\lambda^{2}\;\rightarrow\;0
  \;\;\mbox{if}\;\;m_{1}=m_{2}\;\rightarrow\;\infty
\end{equation}
Therefore, in the limit \mbox{$m_{1}=m_{2}\;\rightarrow\;\infty$}
on the triviality surface, [$M_{n}^{2}$] and [$M_{c}^{2}$] can be
solved up to order $O(h^{2})$ exactly:
\begin{eqnarray}
  \mbox{Eigenvalues of}\; [M_{n}^{2}]\,&=&\,
[2m_{1}^{2}+O(h^{2}),\; 2m_{1}^{2}+O(h^{2}),\; O(h^{2}),\;
                 O(h^{2}),\; O(h^{2}),\; 0] \nonumber  \\
  \mbox{Eigenvalues of}\; [M_{c}^{2}]\,&=&\,[2m_{1}^{2}+O(h^{2}),\;
2m_{1}^{2}+O(h^{2}),\; 0,\; 0]
\end{eqnarray}
The square roots of the seven non-zero eigenvalues in (39) are just the
seven triviality upper bounds on the seven Higgs masses respectively,
including the triviality upper bound on $m_{LH}$.
(39) together with (38) explains why $B_{HH}\,=\,\sqrt{2}\,m_{1}$ is
valid up to order
$O(h^{2})$ when $m_{1}^{2}$ is large.
(39) also implies that there are exactly
three light neutral Higgs bosons
when $m_{1}^{2}$ is large. As for the
lightest-Higgs mass $m_{LH}$, (39) indicates that the triviality
upper bound
on $m_{LH}$ is of order $O(h)$,
which is consistent with the upper bound $B_{LH}^{(BS)}$:
\mbox{$m_{LH}\leq B_{LH}^{(BS)}=\frac{h}{\sqrt{2}}v$} in (37).
Due to (38), the triviality upper bound on $m_{LH}$
decreases to zero monotonically as \mbox{$m_{1}=m_{2}\,\rightarrow\,
\infty$}.
According to the present experimental lower bound on the Higgs mass
\cite{ref26,ref27}, we require that the
triviality upper bound on $m_{LH}$ be larger than $1\;M_{W}$, and
this requirement leads to an upper bound on $m_{1}\,(=m_{2})$ due to the
fact that the triviality upper bound on $m_{LH}$ decreases to zero
as \mbox{$m_{1}=m_{2}\,\rightarrow\,\infty$}.
An explicit realization of this idea is given in Fig.2, where
$(h,\;\lambda)=(h_{A},\;\lambda_{A})$, $\phi\,=\,0.6$, and
$({\tilde{v}}_{3},\;m_{1})$ is examined thoroughly.
The enclosed region of Fig.2
is the allowed range of ${\tilde{v}}_{3}$ versus $m_{1}$.
An interesting
quantity $B_{soft}$ can be defined in such a way that
the allowed range of
${\tilde{v}}_{3}$ shrinks to a single point at \mbox{$m_{1}=B_{soft}$}
and there is no solution for \mbox{$m_{1}>B_{soft}$}. In Fig.2,
$B_{soft}\,=\,138\;M_{W}$. The meaning of $B_{soft}$ is clear: For given
$\phi$, $B_{soft}$ is the absolute upper bound on $m_{1}$ for
\mbox{$0\,<\,{\tilde{v}}_{3}\,\leq\,\frac{M_{W}}{2g_{2}}$}.
That is, $B_{soft}$ is the
upper bound on $m_{1}$ when ${\tilde{v}}_{3}\,=\,\frac{M_{W}}{2g_{2}}$,
and the
upper bound on $m_{1}$ is smaller than $B_{soft}$ when
${\tilde{v}}_{3}\,<\,\frac{M_{W}}{2g_{2}}$. In fact, $B_{soft}$
can be interpreted
as the absolute upper bound (with $\phi$ fixed) on all
the soft SUSY-breaking
parameters \mbox{($m_{1},\;m_{2},\;{\tilde{m}}_{12}$)} because
\mbox{$m_{1}^{2}\,\approx\,{\tilde{m}}_{12}^{2}$}
is true on the triviality surface when $m_{1}^{2}$ is large.
The fact that \mbox{$m_{1}^{2}\,\approx\,{\tilde{m}}_{12}^{2}$}
on the triviality surface when $m_{1}^{2}$ is large
can be explained by the observations that
$h^{2}\;\propto\;({\tilde{m}}_{12}^{2}-m_{1}^{2})$ and that
$h^{2}$ is negligible when $m_{1}^{2}$ is large.

The dependence of $B_{soft}$ on $\phi$ is displayed in Fig.3, where
$(h_{A},\;\lambda_{A})$ and $(h_{B},\;\lambda_{B})$
have identical results.
In Fig.3, it is also required that the triviality upper bound on
$m_{LH}$ be larger than $1\;M_{W}$.
Fig.3 and Fig.2 form the complete picture of the triviality
upper bound on the soft SUSY-breaking parameters. For example,
$B_{soft}\,=\,2380\;M_{W}$ at $\phi\,=\,0$, and
$B_{soft}\,=\,84.6\;M_{W}$ at $\phi\,=\,\frac{{\pi}}{2}$.
Furthermore, all the
conclusions about $B_{soft}$ can be re-interpreted as the absolute
triviality upper bound on the heaviest-Higgs mass (with $\phi$ fixed)
by means
of \mbox{$B_{HH}\,\simeq\,\sqrt{2}\,B_{soft}$}. Therefore, Fig.3 also
provides the complete picture of the absolute
triviality upper bound on the heaviest-Higgs mass $m_{HH}$.
The computations of Fig.3 are very sensitive to the experimental
lower bound on the Higgs mass. Fig.3 is obtained by
requiring that the triviality upper bound on $m_{LH}$ be larger
than $1\;M_{W}$,
and $B_{soft}\,=\,2380\;M_{W}$ is obtained when $\phi\,=\,0$. However,
$B_{soft}\,=\,4.88\;M_{W}$ at $\phi\,=\,0$ will be obtained if we require
that the triviality upper bound on $m_{LH}$ be larger than $2\;M_{W}$.

Inspired by Fig.2, we can define the absolute triviality lower bound
$B_{N}$ on ${\tilde{v}}_{3}$ for given $\phi$. For example,
$B_{N}\,=\,0.7\;M_{W}$ in Fig.2. $B_{N}$ gives a modest measure of the
constraint on ${\tilde{v}}_{3}$. Fig.4 displays the dependence of
$B_{N}$ on $\phi$ for $(h_{A},\;\lambda_{A})$ and
$(h_{B},\;\lambda_{B})$.
Besides, it is required that the triviality upper bound on
$m_{LH}$ be larger than $1\;M_{W}$.
The dotted straight line corresponds to the absolute upper bound of
$0.749\;M_{W}$ on ${\tilde{v}}_{3}$, (28). For $\phi\,=\,0$
(i.e., no CP-violation in the scalar sector), we have
$0.24\,M_{W}\,\leq\,{\tilde{v}}_{3}\,\leq\,0.749\,M_{W}$.
For $\phi\,=\,\frac{{\pi}}{2}$,
$0.65\,M_{W}\,\leq\,{\tilde{v}}_{3}\,\leq\,0.749\,M_{W}$.
Therefore, triviality is very helpful for a better understanding of
${\tilde{v}}_{3}$. In addition, if the experimental lower bound on
the Higgs
mass is raised in the future, all the bounds involved in Fig.3
and Fig.4 will
become stronger, which implies a better understanding of
the heaviest-Higgs
mass, the VEV of the Higgs singlet, and the soft SUSY-breaking
parameters. However, NMSSM is not consistent with an unlimited raise of
the experimental lower bound on the Higgs mass. In Section 5,
we will derive an
absolute upper bound of $2.8\;M_{W}$ on the lightest-Higgs mass.

Finally, let's check the assumptions of negligible $f_{t}$ and $g_{i}$.
With
the help of \cite{ref15}, all the computations involved in
\mbox{Figures 1-4}
do satisfy the assumption of negligible $g_{i}$.
To check the assumption of negligible $f_{t}$, take
the mass of top quark $m_{t}\,=\,170\;\mbox{GeV}$.
Generally speaking, this assumption is
reasonable when $m_{1}^{2}$ is small,
but it needs modifications when $m_{1}^{2}$ is large
because $h^{2}$ and $\lambda^{2}$ are small according
to (38). As for Fig.4, the computation
of $B_{N}$ indicates that $f_{t}$ is
negligible. However, the
computation of $B_{soft}$
in Fig.3 indicates that $f_{t}^{2}$ is as important as
$h^{2}$ and $\lambda^{2}$. To understand the effect of
$f_{t}^{2}$ on
$B_{soft}$, refer to (6)-(8).
Because all the coefficients of $f_{t}^{2}$-terms
in (6)-(8) are positive (assuming negligible $g_{i}$),
the inclusion of
$f_{t}$ makes triviality even stronger. That is, the Landau pole
$\mbox{t}_{L}$ will be smaller if $f_{t}$ is included. Qualitatively,
it implies that $B_{soft}$ should be smaller (i.e., a
stronger upper bound) if $f_{t}$ is included. In general, all the
triviality bounds will become stronger if $f_{t}$ is included. In other
words, the results of Fig.3 should be regarded as a weak absolute
upper bound on the soft SUSY-breaking parameters
and the heaviest-Higgs mass.
\section{Absolute Upper Bound on the Lightest-Higgs Mass}

\hspace{0.8cm}With the inclusion of the Higgs singlet in NMSSM,
the tree-level
upper bound on the lightest-Higgs mass of MSSM is no
longer valid. Therefore, it is of considerable importance
to study the
triviality upper bound on the lightest-Higgs mass in order to
devise effective
search strategies for the detection of Higgs particles.
For given $\phi$, we can
define a new quantity $B_{LH}$, the absolute triviality
upper bound $B_{LH}$
on the lightest-Higgs mass, as the largest triviality upper bound on
the lightest-Higgs mass with respect to
all the possible values of $m_{1}\,(=m_{2})$ and ${\tilde{v}}_{3}$.
However, (39) implies that a search in the small-$m_{1}^{2}$ regime
is enough, and the
dependence of $B_{LH}$ on $\phi$ is displayed in Fig.5, where
\mbox{line A} and \mbox{line B} correspond to $(h_{A},\;\lambda_{A})$ and
$(h_{B},\;\lambda_{B})$ respectively. It is verified that Fig.5 satisfies
the assumptions of negligible $f_{t}$ and $g_{i}$.

When $\phi\,=\,0$ (i.e., no CP-violation in the scalar sector), the
absolute upper bound \mbox{$B_{LH}\,=\,2.8\;M_{W}$} for \mbox{line B}.
When
$\phi\,=\,\frac{\pi}{2}$, \mbox{$B_{LH}\,=\,1.75\;M_{W}$} for
\mbox{line B}.
Therefore, the absolute triviality upper bound on the lightest-Higgs mass
does lie outside the range of LEP.
\section{Conclusion}

\hspace{0.8cm}With a complete study of the triviality surface,
we are able to derive the triviality bounds on
the heaviest-Higgs mass, the soft SUSY-breaking parameters,
the VEV of the Higgs singlet in \mbox{Section 4},
and the absolute upper bound on the lightest-Higgs mass
in \mbox{Section 5}. Essentially, all the triviality bounds are
derived based on the observations (38) and (39), where
the triviality upper bound on the lightest-Higgs mass
decreases to zero as \mbox{$m_{1}=m_{2}\,\rightarrow\,\infty$}.

The particular choice of the soft SUSY-breaking potential $V_{soft}$
in (2) can be viewed as an unsatisfactory feature
of the present formulation.
However, the triviality bounds derived in \mbox{Sections 4 and 5} persist
even if a more general $V_{soft}$ is
considered. We begin the argument of the above statement
with the most general $V_{soft}$ \cite{ref10,ref21}:
\begin{eqnarray}
  V_{soft}&=&m_{1}^{2}{{\Phi^{\dagger}}_{1}}
{\Phi_{1}}+m_{2}^{2}{{\Phi^{\dagger}}_{2}}{\Phi_{2}}
             -m_{12}^{2}{{\Phi^{\dagger}}_{1}}
{\Phi_{2}}-{m^{\ast}}_{12}^{2}{{\Phi^{\dagger}}_{2}}{\Phi_{1}}
\nonumber \\
          & &+m_{4}^{2}N^{\ast}N+m_{5}^{2}N^{2}+
{m^{\ast}}_{5}^{2}{N^{\ast}}^{2} \nonumber \\
          & &+h\,m_{3}(A_{1}\Phi^{\dagger}_{1}\Phi_{2}N+
{A^{\ast}}_{1}\Phi^{\dagger}_{2}\Phi_{1}{N^{\ast}}) \nonumber \\
          &  & +\frac{1}{3}\,\lambda\,m_{3}(A_{2}N^{3}+
{A^{\ast}}_{2}{N^{\ast}}^{3})
\end{eqnarray}
Now, the relevant parameter space consists of:
\begin{equation}
  (h,\;\lambda,\;v_{1},\;v_{2},\;v_{3},\;m_{1}^{2},\;m_{2}^{2},\;
m_{3},\;m_{4}^{2},\;m_{5}^{2},\;m_{12}^{2},\;A_{1},\;A_{2})
\end{equation}
plus three complex vacuum constraints on the parameters derived from
the minimization of $V'=V+V_{soft}$ and (19). Choosing
$\tan\beta=1$, we have $m_{1}^{2}=m_{2}^{2}$ from
the minimization of $V'$ again. In the limit of large
$\Lambda_{L}\,(=B_{HH})$, the Landau pole (13) always implies:
\begin{equation}
  h^{2}\,\rightarrow\,0\;\;\mbox{if}\;\;\Lambda_{L}\,(=B_{HH})\,
\rightarrow\,\infty
\end{equation}
In the large-$m_{HH}$ limit (e.g., in the large-$m_{1}$ limit)
on the triviality surface, (42) implies that $[M_{n}^{2}]$
and $[M_{c}^{2}]$ can be solved up to order $O(h)$ exactly:
\begin{eqnarray}
  \mbox{Eigenvalues of}\; [M_{n}^{2}]\,&=&\,[2m_{1}^{2}+O(h),\;
2m_{1}^{2}+O(h),
        \;m_{(+)}+O(h),\;m_{(-)}+O(h),\; O(h),\; 0] \nonumber  \\
  \mbox{Eigenvalues of}\; [M_{c}^{2}]\,&=&\,[2m_{1}^{2}+O(h),\;
2m_{1}^{2}+O(h),\; 0,\; 0] \nonumber \\
  m_{(\pm)}&=&m_{4}^{2}+4\lambda^{2}{|v_{3}|}^{2}\pm
              2|m_{5}^{2}+{\lambda}m_{3}v_{3}A_{2}+
{\lambda^{2}}{v_{3}^{\ast}}^{2}|
\end{eqnarray}
With the most general $V_{soft}$ (40), there is, in general, exactly
one Higgs boson staying light in the large-$m_{HH}$ limit. According to
(42) and (43), the triviality upper bound on $m_{LH}$ is of order
$O(h^{\frac{1}{2}})$, and decreases to zero as
\mbox{$m_{HH}\,\rightarrow\,\infty$}.
This observation implies that the analyses of
\mbox{Sections 4 and 5} still apply to the
most general $V'=V+V_{soft}$. That is, the triviality bounds
on the heaviest-Higgs mass, the lightest-Higgs mass, the VEV of the
Higgs singlet, and the soft SUSY-breaking parameters will not be
lost even if the largest parameter space of NMSSM is considered.

Finally, two aspects of the present computations can be improved.
First, the computation of the effective potential in this paper is
performed only at tree level. Because the contributions of the top
and stop loops are important, it is necessary for future works
to include one-loop contributions. Second, $\tan\beta =1$ is chosen
in this paper for the sake of simplicity. However, this particular
choice has no physical motivation. Therefore, choices different
from $\tan\beta =1$ should be considered and the discussion of the
$\tan\beta$-dependence may be a point of interest in future works.
\section*{Acknowledgement}

\hspace{0.8cm}I would like to thank Professor Mary K. Gaillard
for her support and nice advice. I also thank \mbox{Dr. H.-C. Cheng}
for discussions about MSSM.
This work was supported in part by the Director, Office of
Energy Research, Office of High Energy and Nuclear Physics, Division of
High Energy Physics of the U.S. Department of Energy under Contract
DE-AC03-76SF00098 and in part by the National Science Foundation under
grant PHY90-21139.
\pagebreak

\pagebreak
\clearpage
\hspace{2in} FIGURE CAPTIONS
\vskip 0.5in
Fig.1: A plot of the triviality upper bound $B_{HH}$
($M_{W}$) on the heaviest-Higgs
mass versus $m_{1}^{2}$ ($M_{W}^{2}$) for $\phi =0$,
${\tilde{v}}_{3}=0.7\,M_{W}$, and
$(h,\;\lambda)=(h_{A},\;\lambda_{A})$, where the
unit \mbox{$M_{W}=80$ GeV}.
\vskip 0.5in
Fig.2: A plot of the allowed range (the enclosed region)
of ${\tilde{v}}_{3}$ ($M_{W}$)  versus $m_{1}$ ($M_{W}$)
for $\phi =0.6$,
$(h,\;\lambda)=(h_{A},\;\lambda_{A})$, where the unit
\mbox{$M_{W}=80$ GeV}.
The allowed range of ${\tilde{v}}_{3}$
shrinks to a point at
$m_{1}$=\mbox{138 GeV}$\equiv B_{soft}$.
\vskip 0.5in
Fig.3: The plot of the absolute triviality
upper bound $B_{soft}$
($M_{W}$) versus $\phi$ (the unit
\mbox{$M_{W}=80$ GeV}),
where the two solutions $(h_{A},\;\lambda_{A})$ and
$(h_{B},\;\lambda_{B})$ have identical results.
$B_{soft}$ is periodic in
$\phi$ with period $\pi$.
\vskip 0.5in
Fig.4: The plot of $B_{N}$ ($M_{W}$), the
absolute triviality lower bound on
${\tilde{v}}_{3}$, versus $\phi$ (the unit
\mbox{$M_{W}=80$ GeV}),
where the dashed line corresponds to
$(h_{A},\;\lambda_{A})$ and the
solid line corresponds to
$(h_{B},\;\lambda_{B})$. $B_{N}$ is periodic
in $\phi$ with period
$\pi$. The dotted line corresponds to the
absolute upper bound of
$\frac{M_{W}}{2g_{2}}$ on ${\tilde{v}}_{3}$.
\vskip 0.5in
Fig.5: The plot of $B_{LH}$ ($M_{W}$), the
absolute triviality upper bound on the
lightest-Higgs mass versus $\phi$ (the unit
\mbox{$M_{W}=80$ GeV}),
where the dashed line corresponds to
$(h_{A},\;\lambda_{A})$ and the
solid line corresponds to
$(h_{B},\;\lambda_{B})$. $B_{LH}$
is periodic in $\phi$ with period $\pi$.
\end{document}